\documentclass[aps,prd,preprint,superscriptaddress,showpacs,floatfix,nobibnotes]{revtex4}

\usepackage{latexsym}
\usepackage{amsmath}
\usepackage{amssymb}
\usepackage{graphicx}
\usepackage{longtable}

\usepackage{bm}
\usepackage{epsfig}
\usepackage{citesort}

\begin{document}
\title{Fluctuations and Correlations of Conserved Charges near the QCD Critical Point}

\author{Wei-jie Fu}
\email[]{wjfu@itp.ac.cn} \affiliation{Kavli Institute for
Theoretical Physics China (KITPC), Key Laboratory of Frontiers in
Theoretical Physics, Institute of Theoretical Physics, Chinese
Academy of Science, Beijing 100190, China}

\author{Yue-liang Wu}
\email[]{ylwu@itp.ac.cn} \affiliation{Kavli Institute for
Theoretical Physics China (KITPC), Key Laboratory of Frontiers in
Theoretical Physics, Institute of Theoretical Physics, Chinese
Academy of Science, Beijing 100190, China}

\date{\today}

\begin{abstract}
We study the fluctuations and correlations of conserved charges,
such as the baryon number, the electric charge and the strangeness,
at the finite temperature and nonzero baryon chemical potential in
an effective model. The fluctuations are calculated up to the
fourth-order and the correlations to the third-order. We find that
the second-order fluctuations and correlations have a peak or valley
structure when the chiral phase transition takes place with the
increase of the baryon chemical potential; the third-order
fluctuations and correlations change their signs during the chiral
phase transition and the fourth-order fluctuations have two maximum
and one minimum. We also depict contour plots of various
fluctuations and correlations of conserved charges in the plane of
temperature and baryon chemical potential. It is found that higher
order fluctuations and correlations of conserved charges are
superior to the second-order ones to be used to search for the
critical point in heavy ion collision experiments.
\end{abstract}


\maketitle

\section{Introduction}
\vspace{5pt}

Studies of QCD thermodynamics and QCD phase diagram have attracted
lots of attentions in recent years. People believe that deconfined
quark gluon plasma (QGP) is formed in ultrarelativistic heavy ion
collisions~\cite{Shuryak2004,Gyulassy2005,Shuryak2005,Arsene2005,Back2005,Adams2005,Adcox2005,Blaizot2007}.
Various field theory models
studies~\cite{Asakawa1989,Barducci1989,Barducci1994,Berges1999,Halasz1998,Scavenius2001,Hatta2003,Barducci2005,Fu2008}
indicate that there is a critical point in the QCD phase diagram in
the plane of temperature and baryon chemical
potential~\cite{Stephanov2006}, which separates the first-order
phase transition at high baryon chemical potential from the
continuous crossover at high temperature. Although there is no
definite evidence that the QCD critical point also exists in the
lattice QCD calculations, due to the sign problem at finite chemical
potential, some lattice groups find that the QCD critical point
maybe exist in the phase diagram, based on extrapolating results at
small $\mu/T$ (ratio of the chemical potential and the temperature)
to those at large $\mu/T$~\cite{Fodor2002,Ejiri2004,Gavai2005}. In
the meantime, experiments with the goal to search for the QCD
critical point are planed and underway at the Relativistic Heavy Ion
Collider (RHIC) at the Brookhaven National Laboratory (BNL) and at
the Super Proton Synchrotron (SPS) at CERN in
Geneva~\cite{Mohanty2009,Anticic2009,Aggarwal2010}

In order to map the QCD phase diagram, locating the QCD critical
point is a crucial and vital task. It has been known that the
event-by-event fluctuations of various particle multiplicities are
enhanced in heavy ion collisions that freeze out near the critical
point. Therefore the QCD critical point can be found through the
non-monotonic behavior of various fluctuation observables as a
function of varying control
parameters~\cite{Asakawa2000,Jeon2000,Stephanov1998,Stephanov1999,Hatta2003b,Jeon2004,Stephanov2009,Asakawa2009,Stephanov2010,Athanasiou2010}.
Particularly, fluctuations of conserved charges, such as the baryon
number, electric charge, and strangeness, deserve more attentions.
On the one hand, the fluctuations of conserved charges are sensitive
to the structure of the thermal strongly interacting matter and
behave differently between the hadronic and QGP
phases~\cite{Asakawa2000,Jeon2000,Hatta2003,Jeon2004,Bhattacharyya2010}.
On the other hand, since the conserved charges are conserved through
the evolution of the fire ball, the fluctuations of conserved
charges can be measured in heavy ion collision experiments.

In our previous work~\cite{Fu2010}, we have studied the fluctuations
and correlations of conserved charges in the 2+1 flavor
Polyakov--Nambu--Jona-Lasinio (PNJL) model at finite temperature and
zero chemical potential. We made an interesting comparison with the
recent lattice calculations which were performed with an improved
staggered fermion action at two values of the lattice cutoff with
almost physical up and down quark masses and a physical value for
the strange quark mass~\cite{Cheng2009}. It has been seen that our
calculated results are well consistent with those obtained in
lattice calculations, which indicates that the 2+1 flavor PNJL model
is well applicable to study the fluctuations and correlations of
conserved charges. The validity of this effective model is expected,
since the critical behavior of the QCD phase transition is governed
by the universality class of the chiral symmetry, which is kept in
this model. Furthermore, compared with the conventional
Nambu--Jona-Lasinio model, the PNJL model not only has the chiral
symmetry and its dynamical breaking mechanism, but also includes the
effect of color confinement through the Polyakov
loop~\cite{Meisinger9602,Pisarski2000,Fukushima2004,Ratti2006a,Ratti2006b,Ciminale2008,Zhang2006,Fu2008,Fu2009,Ghosh2006}.

In this work, we will extend our previous work~\cite{Fu2010} to the
cases with nonzero baryon chemical potential. Such an extension is
quite nontrivial, which allows us to use the fluctuations and
correlations of conserved charges to explore the critical behavior
of the QCD critical point and to find out the location of the
critical point. Furthermore, when the baryon chemical potential does
not vanish, the fluctuations and correlations of odd order develop
finite values, which are identical to zero at vanishing baryon
chemical potential. In this paper we shall pay our most attentions
on studies of the non-monotonic behavior of the fluctuations and
correlations of conserved charges near the QCD critical point and
depict the contour plots of various fluctuations and correlations in
the plane of temperature and baryon chemical potential.

This paper is organized as follows. In Sec. II we introduce the
fluctuations and correlations of conserved charges and simply review
the formalism of the 2+1 flavor PNJL model. In Sec. III we give our
calculated numerical results of the fluctuations of conserved
charges. In Sec. IV we present the numerical results of the
correlations among conserved charges. Our summary and conclusions
are given in Sec. V.

\section{Fluctuations and Correlations}

In this section we focus on the cumulants of the conserved charge
multiplicity distributions, which can be expressed as the derivative
of the pressure ($P$) of a thermodynamical system with respective to
its chemical potentials corresponding to conserved charges, i.e.
\begin{equation}
\chi_{ijk}^{BQS}=\frac{\partial^{i+j+k}(P/T^{4})}
{\partial(\mu_{B}/T)^{i}\partial(\mu_{Q}/T)^{j}\partial(\mu_{S}/T)^{k}},\label{susceptibility}
\end{equation}
where $T$ is the temperature, and $\mu_{B,Q,S}$ are the chemical
potentials for baryon number, electric charge, and strangeness,
respectively. They are related with the quark chemical potentials
through the following relations,
\begin{equation}
\mu_{u}=\frac{1}{3}\mu_{B}+\frac{2}{3}\mu_{Q},\quad
\mu_{d}=\frac{1}{3}\mu_{B}-\frac{1}{3}\mu_{Q},\quad\textrm{and}\quad
\mu_{s}=\frac{1}{3}\mu_{B}-\frac{1}{3}\mu_{Q}-\mu_{S}.\label{chemicalpotential}
\end{equation}
where $\mu_{u,d,s}$ are the chemical potentials for $u$, $d$, and
$s$ quarks, respectively. Denoting the ensemble average of conserved
charge number $N_{X}$ ($X=B,Q,S$) with $\langle N_{X}\rangle$, we
can obtain the second and higher order fluctuations of the conserved
charges as follow
\begin{eqnarray}
\chi_{2}^{X}&=&\frac{1}{VT^{3}}\langle {\delta N_{X}}^{2} \rangle,\label{chi2}\\
\chi_{3}^{X}&=&\frac{1}{VT^{3}}\langle {\delta N_{X}}^{3} \rangle,\label{chi3}\\
\chi_{4}^{X}&=&\frac{1}{VT^{3}}\Big(\langle {\delta
N_{X}}^{4}\rangle-3{\langle {\delta N_{X}}^{2}
\rangle}^{2}\Big),\label{chi4}
\end{eqnarray}
where $\delta N_{X}\equiv N_{X}-\langle N_{X}\rangle$ and $V$ is the
volume of the system. In the same way, we can also obtain mixed
cumulants of the conserved charge distributions, which are known as
the correlations among conserved charges. For example,
\begin{eqnarray}
\chi_{11}^{XY}&=&\frac{1}{VT^{3}}\langle \delta N_{X} \delta N_{Y}\rangle,  \\
\chi_{12}^{XY}&=&\frac{1}{VT^{3}}\langle \delta N_{X}{\delta N_{Y}}^{2}\rangle,\\
\chi_{111}^{XYZ}&=&\frac{1}{VT^{3}}\langle \delta N_{X} \delta N_{Y}
\delta N_{Z} \rangle,
\end{eqnarray}
It should be noted that when the chemical potentials are vanishing,
i.e. $\mu_{B,Q,S}=0$, the fluctuations and correlations of conserved
charges in Eq.(\ref{susceptibility}) (also known as generalized
susceptibilities) are nonvanishing only when $i+j+k$ is even, but
when $\mu_{B}$ has finite values, which is the case discussed in
this paper, the generalized susceptibilities also develop finite
values when $i+j+k$ is odd.

We now adopt the 2+1 flavor Polyakov-loop improved NJL model to
study the fluctuations and correlations of conserved charges near
the QCD critical point, which is a nontrivial extension to our
previous work~\cite{Fu2010}, where the fluctuations and correlations
of conserved charges were calculated in the 2+1 flavor PNJL model at
finite temperature but with vanishing chemical potentials. It is
interesting to notice that the calculated results in ~\cite{Fu2010}
are well consistent with those obtained in lattice
calculations~\cite{Cheng2009}, which shows that the 2+1 flavor PNJL
model is well applicable to study the cumulants of conserved charge
multiplicity distributions. Before a detailed study, let us give a
brief review on the 2+1 flavor PNJL model for completeness. Details
about the model can be found in Ref.~\cite{Fu2008}.

The Lagrangian density for the 2+1 flavor PNJL model is given as
\begin{eqnarray}
\mathcal{L}_{\mathrm{PNJL}}&=&\bar{\psi}(i\gamma_{\mu}D^{\mu}+\gamma_{0}
 \hat{\mu}-\hat{m}_{0})\psi
 +G\sum_{a=0}^{8}\Big[(\bar{\psi}\tau_{a}\psi)^{2}
 +(\bar{\psi}i\gamma_{5}\tau_{a}\psi)^{2}\Big]   \nonumber \\
&&-K\Big[\textrm{det}_{f}(\bar{\psi}(1+\gamma_{5})\psi)
 +\textrm{det}_{f}(\bar{\psi}(1-\gamma_{5})\psi)\Big]
 -\mathcal{U}(\Phi,\Phi^{*} \, ,T),\label{lagragian}
\end{eqnarray}
where $\psi=(\psi_{u},\psi_{d},\psi_{s})^{T}$ is the three-flavor
quark field, and
\begin{equation}
D^{\mu}=\partial^{\mu}-iA^{\mu}\quad\textrm{with}\quad
A^{\mu}=\delta^{\mu}_{0}A^{0}\quad\textrm{,}\quad
A^{0}=g\mathcal{A}^{0}_{a}\frac{\lambda_{a}}{2}=-iA_4,
\end{equation}
where $\lambda_{a}$'s are the Gell-Mann matrices in color space and
$g$ is the gauge coupling strength.
$\hat{m}_{0}=\textrm{diag}(m_{0}^{u},m_{0}^{d},m_{0}^{s})$ is the
three-flavor current quark mass matrix. Throughout this work, we
take $m_{0}^{u}=m_{0}^{d}\equiv m_{0}^{l}$, while keep $m_{0}^{s}$
being larger than $m_{0}^{l}$, which breaks the $SU(3)_f$ symmetry.
The matrix $\hat{\mu}=\textrm{diag}(\mu_{u}, \mu_{d}, \mu_{s})$
denotes the quark chemical potentials which are related with the
conserved charge chemical potentials through relations in
Eq.(\ref{chemicalpotential}).

In the above PNJL Lagrangian,
$\mathcal{U}\left(\Phi,\Phi^{*},T\right)$ is the Polyakov-loop
effective potential, which is expressed in terms of the traced
Polyakov-loop $\Phi=(\mathrm{Tr}_{c}L)/N_{c}$ and its conjugate
$\Phi^{*}=(\mathrm{Tr}_{c}L^{\dag})/N_{c}$ with the Polyakov-loop
$L$ being a matrix in color space given explicitly by
\begin{equation}
L(\vec{x})=\mathcal{P}\exp\Big[i\int_{0}^{\beta}d\tau\,
A_{4}(\vec{x},\tau)\Big] =\exp[i \beta A_{4}],
\end{equation}
with  $\beta=1/T$ being the inverse of temperature and
$A_{4}=iA^{0}$.

In our work, we use the Polyakov-loop effective potential which is a
polynomial in $\Phi$ and $\Phi^{*}$~\cite{Ratti2006a}, given by

\begin{equation}
\frac{\mathcal{U}(\Phi,\Phi^{*},T)}{T^{4}} =
-\frac{b_{2}(T)}{2}\Phi^{*}\Phi -\frac{b_{3}}{6}
(\Phi^{3}+{\Phi^{*}}^{3})+\frac{b_{4}}{4}(\Phi^{*}\Phi)^{2} \, ,
\end{equation}
with
\begin{equation}
b_{2}(T)=a_{0}+a_{1}(\frac{T_{0}}{T})+a_{2} {(\frac{T_{0}}{T})}^{2}
+a_{3}{(\frac{T_{0}}{T})}^{3}.
\end{equation}
The parameters in the effective potential are fitted to reproduce
the thermodynamical behavior of the pure-gauge QCD obtained from the
lattice simulations, and their values are given in
Table~\ref{pol_para}. The parameter $T_{0}$ is the critical
temperature for the deconfinement phase transition to take place in
pure-gauge QCD and $T_{0}$ is chosen to be $270\,\mathrm{MeV}$
according to the lattice calculations.

\begin{table}[!htb]
\begin{center}
\caption{Parameters for the Polyakov-loop effective potential
$\mathcal{U}$} \label{pol_para}
\begin{tabular}{cccccc}
\hline \hline \vspace{0.1cm}
$a_{0}$\qquad\qquad&$a_{1}$\qquad\qquad&$a_{2}$\qquad\qquad&$a_{3}$\qquad\qquad&
$b_{3}$\qquad\qquad& $b_{4}$\\
\hline 6.75 \qquad\qquad& $-1.95$ \qquad\qquad& 2.625\qquad\qquad& $-7.44$\qquad\qquad& 0.75\qquad\qquad& 7.5\\
\hline
\end{tabular}
\end{center}
\end{table}

In the mean field approximation, the thermodynamical potential
density for the 2+1 flavor quark system is given by
\begin{eqnarray}
\Omega&=&-2N_{c}\sum_{f=u,d,s}\int\frac{d^{3}p}{(2\pi)^{3}}\Big\{E_{p}^{f}\theta(\Lambda^{2}-p^{2})\nonumber \\
&&+ \frac{T}{3}\ln\big[1+3\Phi^{*}e^{-(E_{p}^{f}-\mu_{f})/T}+3\Phi
e^{-2(E_{p}^{f}-\mu_{f})/T}+e^{-3(E_{p}^{f}-\mu_{f})/T}\big]
\nonumber \\
&&+ \frac{T}{3}\ln\big[1+3\Phi e^{-(E_{p}^{f}+\mu_{f})/T}+3\Phi^{*}
e^{-2(E_{p}^{f}+\mu_{f})/T}+e^{-3(E_{p}^{f}+\mu_{f})/T}\big]\Big\}\nonumber \\
&&+2G({\phi_{u}}^{2}
+{\phi_{d}}^{2}+{\phi_{s}}^{2})-4K\phi_{u}\,\phi_{d}\,\phi_{s}+\mathcal{U}(\Phi,\Phi^{*},T),\label{thermopotential}
\end{eqnarray}

where $\phi_{i}$'s $(i=u,d,s)$ are the quark chiral condensates, and
the energy-momentum dispersion relation is
$E_{p}^{i}=\sqrt{p^{2}+M_{i}^{2}}$, with the constituent mass being
\begin{equation}
M_{i}=m_{0}^{i}-4G\phi_{i}+2K\phi_{j}\,\phi_{k}.\label{constituentmass}
\end{equation}

Minimizing the thermodynamical potential in
Eq.~\eqref{thermopotential} with respective to $\phi_{u}$,
$\phi_{d}$, $\phi_{s}$, $\Phi$, and $\Phi^{*}$, we obtain a set of
equations for the minimal conditions, which can be solved as
functions of temperature $T$ and three conserved charge chemical
potentials $\mu_{B}$, $\mu_{Q}$, and $\mu_{S}$.

We will use the method of Taylor expansion to compute the
fluctuations and correlations of conserved charges in the PNJL
model. Before numerical calculations, we should fix the five
parameters in the quark sector of the model, whose values used
usually in the literatures are those obtained in
Ref.~\cite{Rehberg1996}, $m_{0}^{l}=5.5\;\mathrm{MeV}$,
$m_{0}^{s}=140.7\;\mathrm{MeV}$, $G\Lambda^{2}=1.835$,
$K\Lambda^{5}=12.36$, and $\Lambda=602.3\;\mathrm{MeV}$, which are
fixed by fitting the observables $m_{\pi}=135.0\;\mathrm{MeV}$,
$m_{K}=497.7\;\mathrm{MeV}$,
$m_{\eta^{\prime}}=957.8\;\mathrm{MeV}$, and
$f_{\pi}=92.4\;\mathrm{MeV}$.

\section{Numerical Results of Fluctuations of Conserved Charges}

\begin{figure}[!htb]
\includegraphics[scale=0.7]{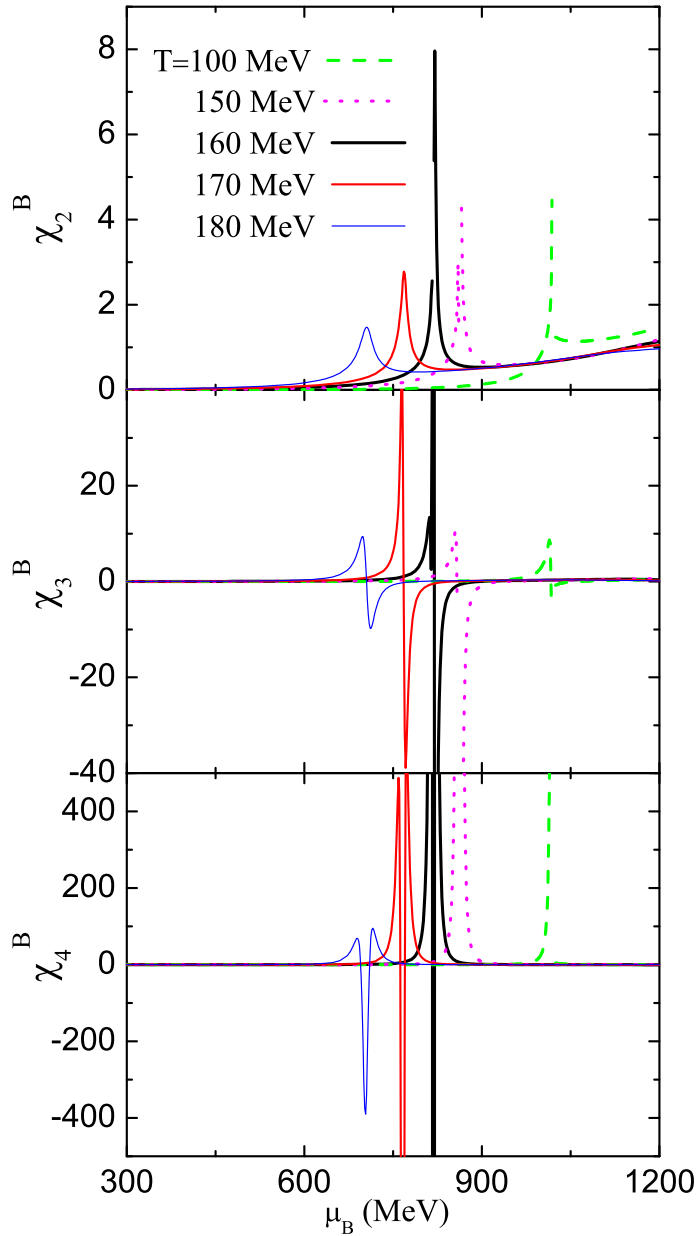}
\caption{(color online). Quadratic (top), cubic (middle), and
quartic (bottom) fluctuations of baryon number as functions of the
baryon chemical potential $\mu_{B}$ ($\mu_{Q}=\mu_{S}=0$) with
several values of temperature in the PNJL model.}\label{f1}
\end{figure}

In this section, we are going to present the numerical results for
the conserved charge fluctuations in the 2+1 flavor PNJL model. It
is found that the QCD critical point is located at about
$T_{c}=160\;\mathrm{MeV}$ and ${\mu_{B}}_{c}=819\;\mathrm{MeV}$
($\mu_{Q}=\mu_{S}=0$) with input parameters given above. This QCD
critical point separates the first-order chiral phase transition at
high baryon chemical potential from the continuous crossover at high
temperature. Our attentions are paid to investigate the behaviors of
the conserved charge fluctuations near the critical point, especial
their singular behaviors, which shed light on the universal symmetry
property of the QCD critical point. Furthermore, the singular
behaviors of conserved charge fluctuations are also very helpful for
searching for the QCD critical point in
experiments~\cite{Athanasiou2010}. In Fig.~\ref{f1} we show the
quadratic, cubic, and quartic fluctuations of baryon number as
functions of $\mu_{B}$ ($\mu_{Q}=\mu_{S}=0$) at several values of
temperature calculated in the PNJL model. From the top panel of
Fig.~\ref{f1}, one can see that $\chi_{2}^{B}$ has a peak structure
when the chiral phase transition takes place, and this peak becomes
sharper and narrower while moving toward the QCD critical point. We
can also clearly notice that the $\chi_{2}^{B}$ diverges at the
critical point, which is explicitly seen from the curve with the
temperature of 160 MeV. According to the definition of cumulants of
the conserved charge multiplicity distributions in
Eq.(\ref{susceptibility}), we have $\chi_{3}^{B}=\partial
\chi_{2}^{B}/\partial(\mu_{B}/T)$. Since $\chi_{2}^{B}$ develops a
cusp during the chiral phase transition, $\chi_{3}^{B}$ changes its
sign there, which is clearly shown in the middle panel of
Fig.~\ref{f1}. This structure of $\chi_{3}^{B}$ was also found by
Asakawa et al.~\cite{Asakawa2009}, who argued that the two sides of
the QCD phase boundary can be distinguished by the sign of
$\chi_{3}^{B}$, therefore the third cumulants carry more information
than the second ones. One can also find that $\chi_{3}^{B}$ diverges
at the QCD critical point. Furthermore, we also calculate the
fourth-order fluctuations of the baryon number, and the results are
shown in the bottom panel of Fig.~\ref{f1}. Based on the above
analysis, it is expected that $\chi_{4}^{B}$ carries even more
information than $\chi_{3}^{B}$, since $\chi_{4}^{B}$ has two
positive maxima and one negative minimum as shown in Fig.~\ref{f1}.
Furthermore, we find that all amplitudes of $\chi_{2}^{B}$,
$\chi_{3}^{B}$, and $\chi_{4}^{B}$ grows rapidly when moving toward
the QCD critical point and finally diverge there.

\begin{figure}[!htb]
\includegraphics[scale=0.8]{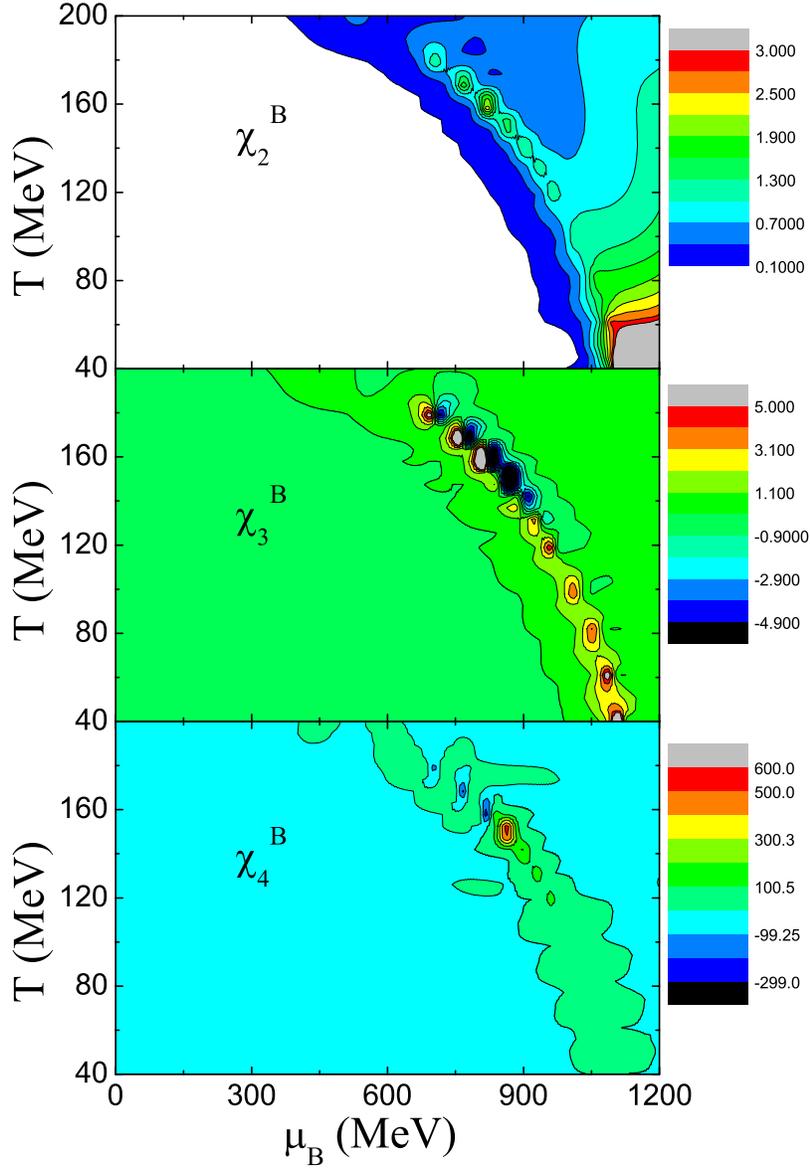}
\caption{(color online). Contour plots of quadratic (top), cubic
(middle), and quartic (bottom) fluctuations of the baryon number as
functions of temperature $T$ and baryon chemical potential $\mu_{B}$
($\mu_{Q}=\mu_{S}=0$) in the PNJL model.}\label{f2}
\end{figure}

In Fig.~\ref{f2} we plot contours of the quadratic, cubic, and
quartic fluctuations of the baryon number as functions of $T$ and
$\mu_{B}$ calculated in the PNJL model. It is clearly seen that the
chiral phase transition line in each of the three contour plots is
obvious, and more important is that the region near around the QCD
critical point also becomes manifest in the three plots, where the
contour lines are dense. Therefore, our calculations demonstrate
that by employing the second and higher order cumulants of baryon
multiplicity distributions, it is possible to search for the QCD
critical point. We should emphasize that in heavy ion collision
experiments finite size and time effects should also be
included~\cite{Athanasiou2010,Stephanov1999}. Comparing higher order
cumulants $\chi_{3}^{B}$ and $\chi_{4}^{B}$ with the quadratic one
$\chi_{2}^{B}$, we observe that the former are superior to the
latter in the search for the QCD critical point. This is because
only when the location is very near the chiral phase transition
line, $\chi_{3}^{B}$ and $\chi_{4}^{B}$ are nonvanishing, while
$\chi_{2}^{B}$ still has finite value when the location is far from
the chiral phase transition line and in the chiral symmetry restored
phase as shown in the top panel of Fig.~\ref{f2}. We should
emphasize that in the top panel of Fig.~\ref{f2}, there is a gray
region in the lower right corner, which is because the quadratic
baryon number fluctuation in this region have relative large value.
However, this region does not correspond to any singular behavior,
since $\chi_{2}^{B}$ in this region is not divergent and it has a
weak dependence on the temperature and baryon chemical potential.
This characteristic is quite different from that in the region near
the QCD critical point, where $\chi_{2}^{B}$ changes rapidly with
$T$ and $\mu_{B}$, and diverges at the QCD critical point, which is
reflected by dense contour lines near the critical point in our
plots.

\begin{figure}[!htb]
\includegraphics[scale=1.2]{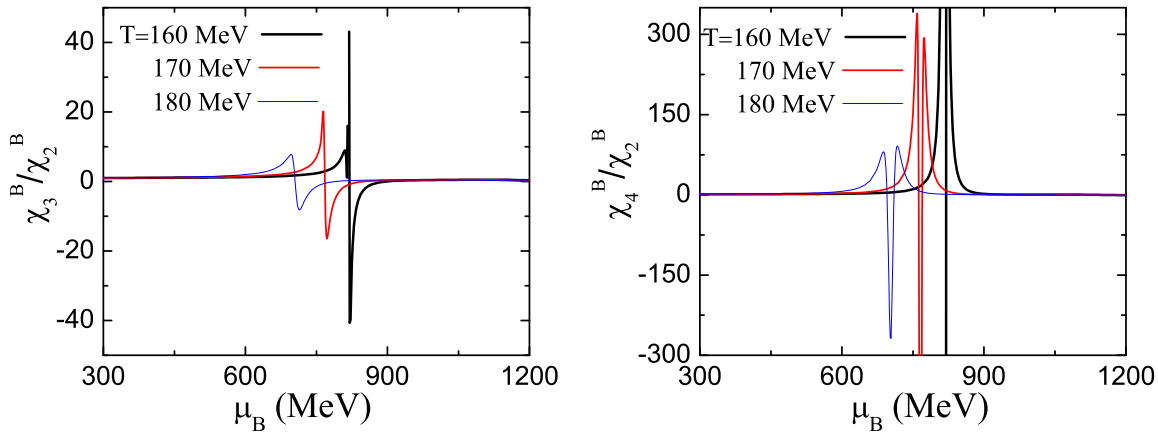}
\caption{(color online). Ratios of cubic to quadratic (left panel)
and those of quartic to quadratic (right panel) baryon number
fluctuations as functions of baryon chemical potential $\mu_{B}$
($\mu_{Q}=\mu_{S}=0$) at several values of temperature in the PNJL
model.}\label{f3}
\end{figure}

In Fig.~\ref{f3} we show $\chi_{3}^{B}/\chi_{2}^{B}$ and
$\chi_{4}^{B}/\chi_{2}^{B}$ versus baryon chemical potential at
several values of temperature calculated in the PNJL model. From
Eq.(\ref{chi2}) to Eq.(\ref{chi4}) we find
\begin{eqnarray}
\frac{\chi_{3}^{B}}{\chi_{2}^{B}}&=&\frac{\langle {\delta N_{B}}^{3} \rangle}{\langle {\delta N_{B}}^{2} \rangle},  \\
\frac{\chi_{4}^{B}}{\chi_{2}^{B}}&=&\frac{\langle {\delta
N_{B}}^{4}\rangle-3{\langle {\delta N_{B}}^{2} \rangle}^{2}}{\langle
{\delta N_{B}}^{2} \rangle}.
\end{eqnarray}
In fact, $\chi_{3}^{B}/\chi_{2}^{B}$ and $\chi_{4}^{B}/\chi_{2}^{B}$
are the skewness and kurtosis of the baryon multiplicity
distributions, respectively, which can be extracted from
event-by-event fluctuations in heavy ion collision experiments.
Furthermore, these ratios are intensive quantities, i.e., being
independent of the volume of the system. In Fig.~\ref{f3} we show
that $\chi_{3}^{B}/\chi_{2}^{B}$ and $\chi_{4}^{B}/\chi_{2}^{B}$
have the same structure as $\chi_{3}^{B}$ and $\chi_{4}^{B}$,
respectively, i.e., $\chi_{3}^{B}/\chi_{2}^{B}$ change its sign
during the chiral phase transition and there are two positive maxima
and one negative minimum on $\chi_{4}^{B}/\chi_{2}^{B}$.
Furthermore, it is observed that the amplitudes of
$\chi_{3}^{B}/\chi_{2}^{B}$ and $\chi_{4}^{B}/\chi_{2}^{B}$ grow
rapidly when moving toward the QCD critical point and become
divergent at the critical point.

\begin{figure}[!htb]
\includegraphics[scale=0.7]{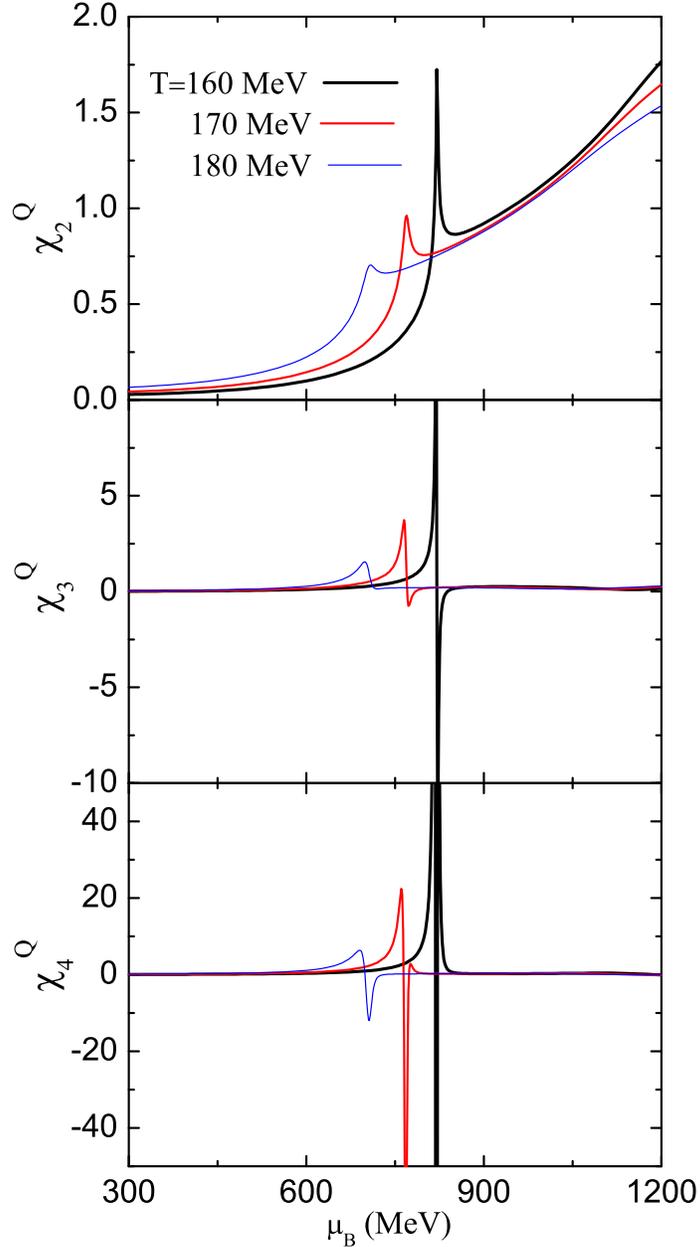}
\caption{(color online). Quadratic (top), cubic (middle), and
quartic (bottom) fluctuations of electric charge as functions of
baryon chemical potential $\mu_{B}$ ($\mu_{Q}=\mu_{S}=0$) with
several values of temperature in the PNJL model.}\label{f4}
\end{figure}

Fig.~\ref{f4} shows the quadratic, cubic, and quartic cumulants of
the electric charge multiplicity distributions as functions of
$\mu_{B}$ calculated in the PNJL model. It is seen that
$\chi_{2}^{Q}$ increases with the baryon chemical potential and has
a cusp during the chiral phase transition. Comparing $\chi_{2}^{Q}$
with $\chi_{2}^{B}$, we observe that the peak in $\chi_{2}^{Q}$
grows less rapidly than that in $\chi_{2}^{B}$. As for higher order
fluctuations of the electric charge, it is found that $\chi_{3}^{Q}$
changes its sign during the chiral phase transition and
$\chi_{4}^{Q}$ has two maxima and one minimum, which are similar to
$\chi_{3}^{B}$ and $\chi_{4}^{B}$, respectively. We should emphasize
that once the location deviates from the chiral phase transition
line, the higher order fluctuations of electric charge approach zero
rapidly, while the second-order fluctuation $\chi_{2}^{Q}$ still has
finite value even the location is far away from the chiral phase
transition line.

\begin{figure}[!htb]
\includegraphics[scale=0.7]{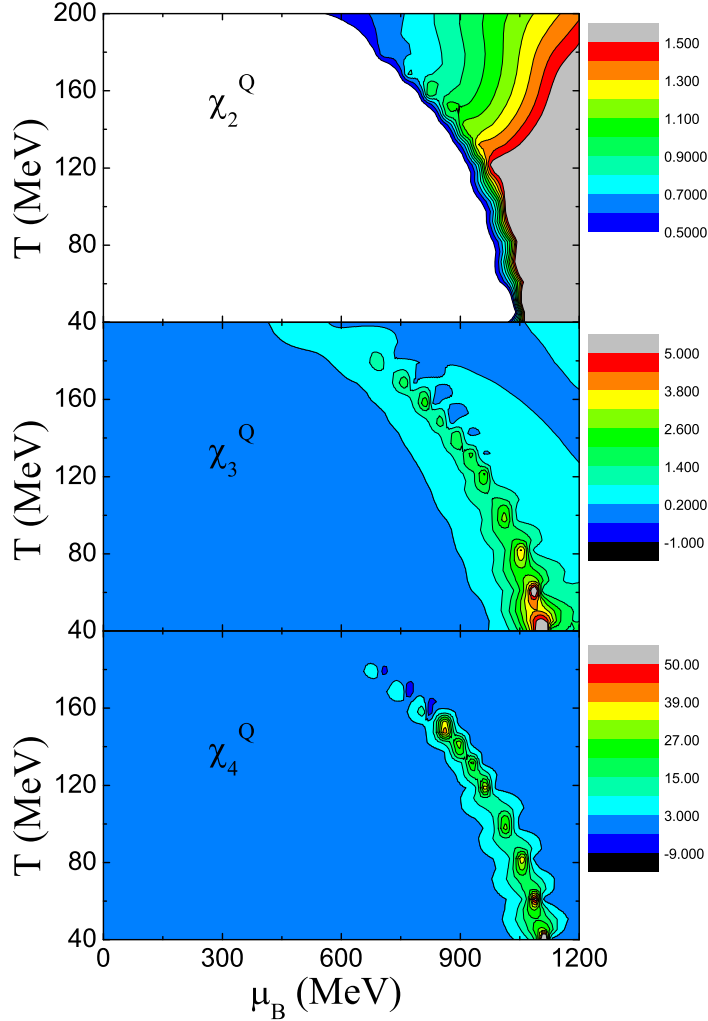}
\caption{(color online). Contour plots of quadratic (top), cubic
(middle), and quartic (bottom) fluctuations of electric charge as
functions of temperature $T$ and baryon chemical potential $\mu_{B}$
($\mu_{Q}=\mu_{S}=0$) in the PNJL model.}\label{f5}
\end{figure}

In Fig.~\ref{f5} we show the contour plots of the quadratic, cubic,
and quartic fluctuations of electric charge as functions of
temperature and baryon chemical potential calculated in the PNJL
model. From the three plots one can easily recognize the chiral
phase transition line, which shows the same features as the contour
plots of baryon number fluctuations given in Fig.~\ref{f2}. However,
we find that employing the quadratic fluctuations of electric charge
to search for the QCD critical point is not easy, since the critical
point in the top panel of Fig.~\ref{f5} is not obvious. While for
the higher order fluctuations of electric charge, it is seen that
the QCD critical point in the contour plots of $\chi_{3}^{Q}$ and
$\chi_{4}^{Q}$ is quite distinct. Therefore, higher order
fluctuations of electric charge are more appropriate for being used
to search for the QCD critical point in heavy ion collision
experiments.

\begin{figure}[!htb]
\includegraphics[scale=0.7]{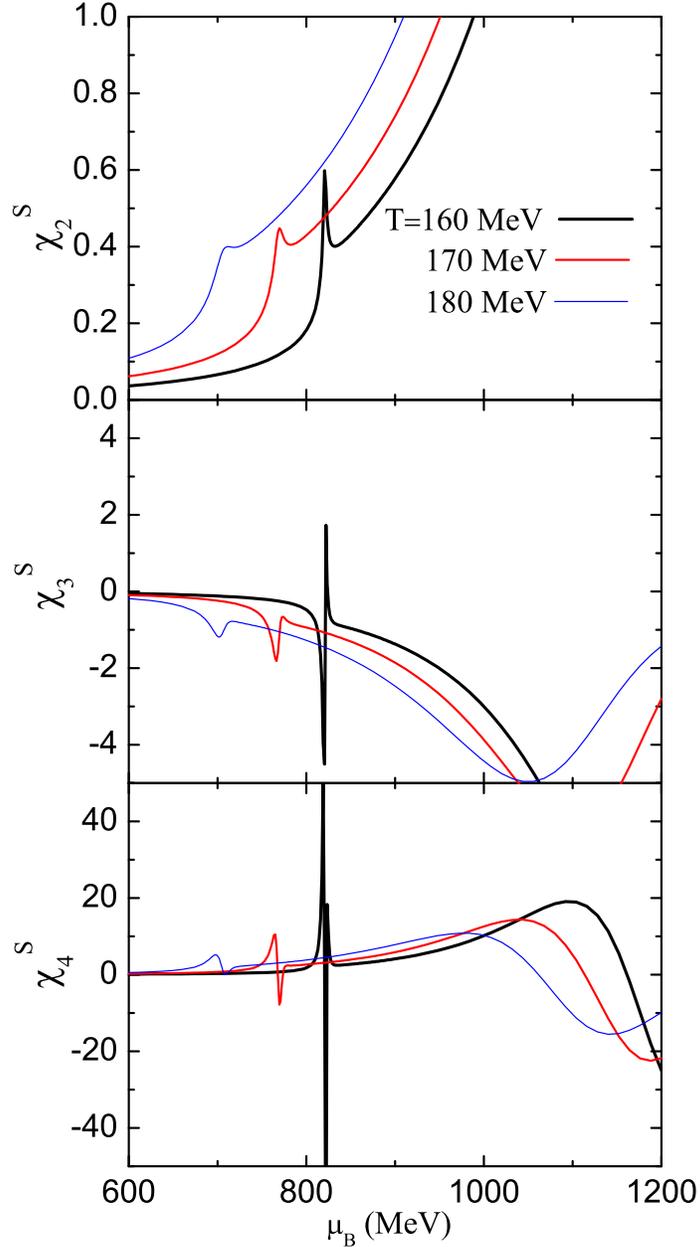}
\caption{(color online). Quadratic (top), cubic (middle), and
quartic (bottom) fluctuations of strangeness as functions of baryon
chemical potential $\mu_{B}$ ($\mu_{Q}=\mu_{S}=0$) with several
values of temperature in the PNJL model.}\label{f6}
\end{figure}

Fig.~\ref{f6} shows the quadratic, cubic, and quartic fluctuations
of strangeness versus the baryon chemical potential at several
values of temperature calculated in the PNJL model. It is noticed
that the second-order and higher order fluctuations of strangeness
are also enhanced when moving toward the QCD critical point, which
are the same as the fluctuations of baryon number and electric
charge. However, contributions to the singularity of the strangeness
fluctuations from the QCD critical point are much less than those to
the singularity of the baryon number or electric charge
fluctuations. In the same way, it is found that higher order
fluctuations of strangeness are superior to the second-order one in
search for the QCD critical point.

\section{Numerical Results of Correlations of Conserved Charges}

\begin{figure}[!htb]
\includegraphics[scale=0.7]{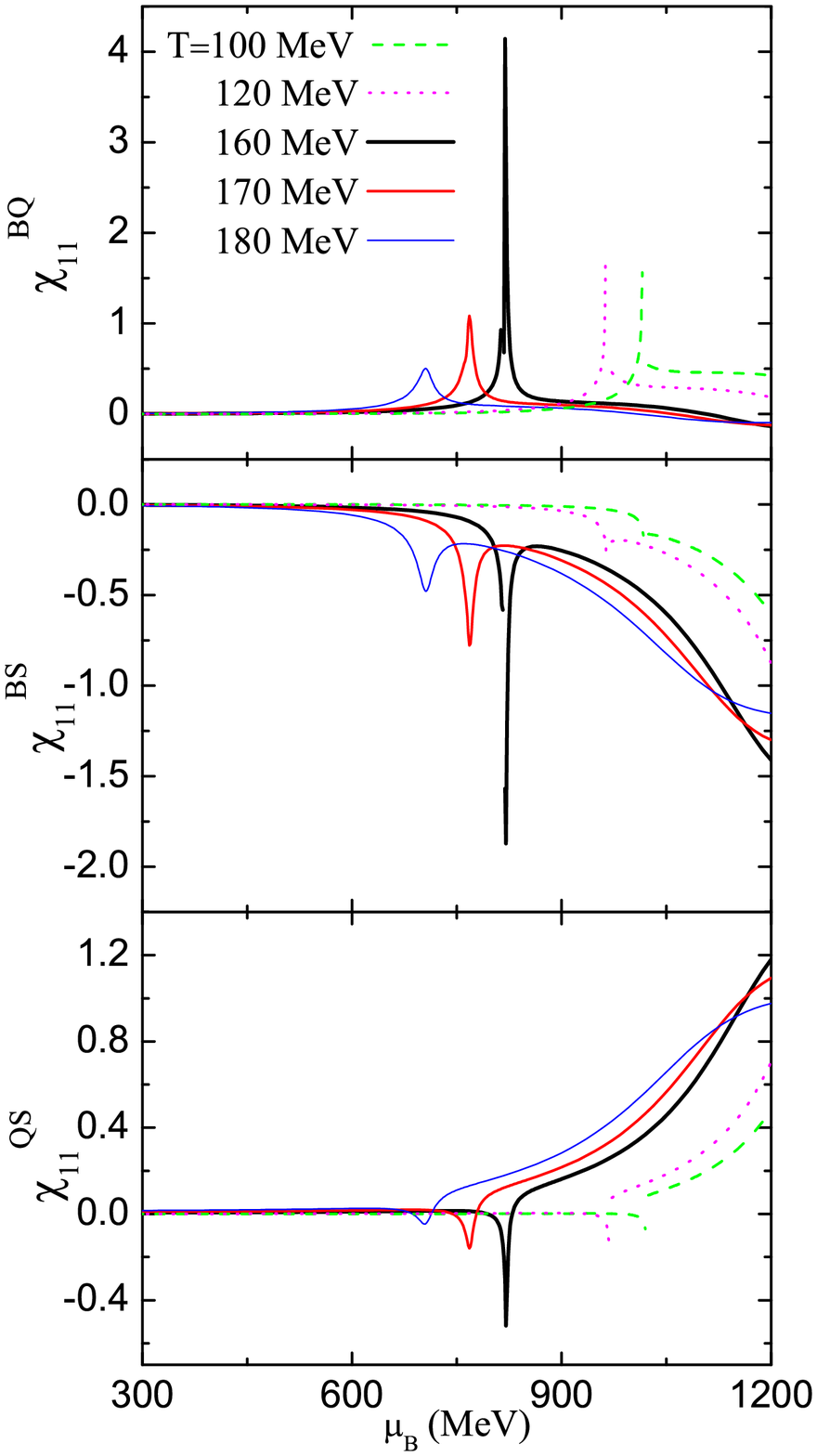}
\caption{(color online). Second-order correlations between baryon
number and electric charge (top), baryon number and strangeness
(middle), electric charge and strangeness (bottom) as functions of
the baryon chemical potential $\mu_{B}$ ($\mu_{Q}=\mu_{S}=0$) with
several values of temperature in the PNJL model.}\label{f7}
\end{figure}

In this section, we are considering the correlations between or
among conserved charges in the 2+1 flavor PNJL model. We will show
that higher order correlations of conserved charges are sensitive to
the critical behaviors related to the QCD critical point and
therefore are very appropriate for being employed to search for the
critical point. In Fig.~\ref{f7} we show the dependence of the
second-order correlations, i.e., $\chi^{BQ}_{11}$, $\chi^{BS}_{11}$,
and $\chi^{QS}_{11}$ on the baryon chemical potential at several
values of temperature in the PNJL model. It is found that all the
second-order correlations have a non-monotonic behavior as functions
of the baryon chemical potential. There is a peak structure on
$\chi^{BQ}_{11}$ during the chiral phase transition;
$\chi^{BS}_{11}$ is negative in the whole chemical potential region
and has a minimum at the phase transition; and $\chi^{QS}_{11}$ also
has a negative minimum but its value is positive in the chiral
symmetry restored phase. Comparing $\chi^{BQ}_{11}$ with
$\chi^{BS}_{11}$ and $\chi^{QS}_{11}$ one can see that
$\chi^{BQ}_{11}$ does not vanish only when the thermodynamical
system is very near the chiral phase transition, while the magnitude
of $\chi^{BS}_{11}$ and $\chi^{QS}_{11}$ increases with the baryon
chemical potential in the chiral symmetric phase as shown in the
middle and bottom panels of Fig.~\ref{f7}. This is because in the
chiral symmetric phase, the system can approximately be described as
noninteracting massless gases, i.e., the Stefan-Boltzmann limit. It
can easily be shown that in the Stefan-Boltzmann limit
$\chi^{BQ}_{11}$ is vanishing while $\chi^{BS}_{11}$ and
$\chi^{QS}_{11}$ have finite values~\cite{Fu2010}.

\begin{figure}[!htb]
\includegraphics[scale=0.7]{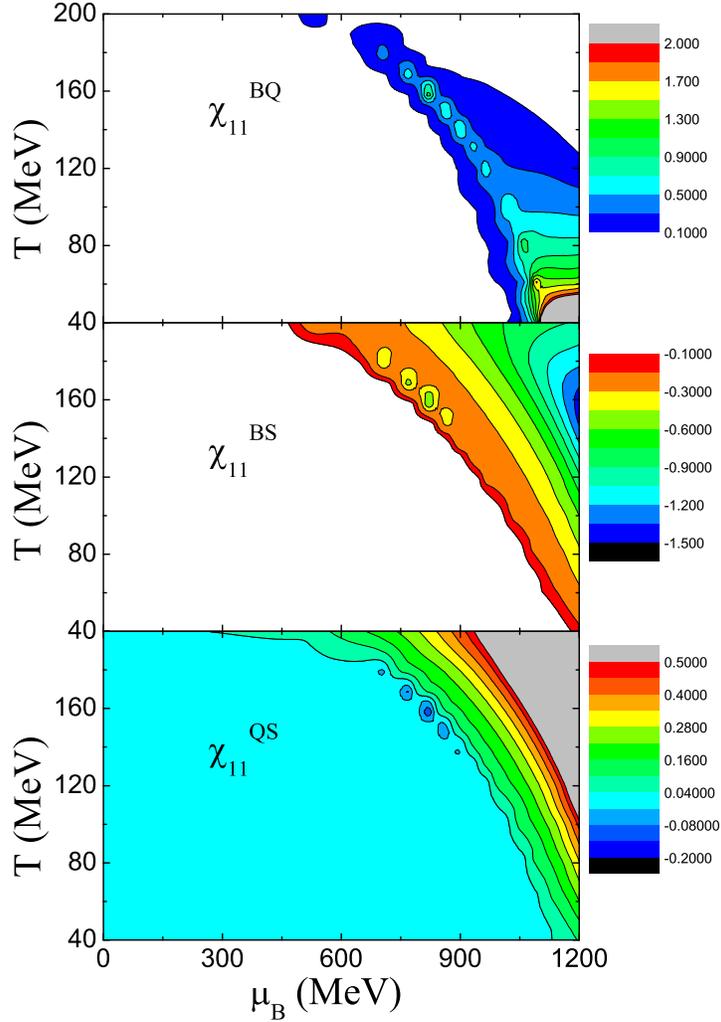}
\caption{(color online). Contour plots of the second-order
correlations $\chi^{BQ}_{11}$ (top), $\chi^{BS}_{11}$ (middle), and
$\chi^{QS}_{11}$ (bottom) as functions of temperature $T$ and baryon
chemical potential $\mu_{B}$ ($\mu_{Q}=\mu_{S}=0$) in the PNJL
model.}\label{f8}
\end{figure}

Fig.~\ref{f8} shows the contour plots of the second-order
correlations $\chi^{BQ}_{11}$, $\chi^{BS}_{11}$, and
$\chi^{QS}_{11}$ as functions of temperature and baryon chemical
potential calculated in the PNJL model. It is found that the chiral
phase transition line in the three plots is distinct, but the QCD
critical point in the plot of $\chi^{BQ}_{11}$ is more apparent than
that in $\chi^{BS}_{11}$ or $\chi^{QS}_{11}$.

\begin{figure}[!htb]
\includegraphics[scale=1.2]{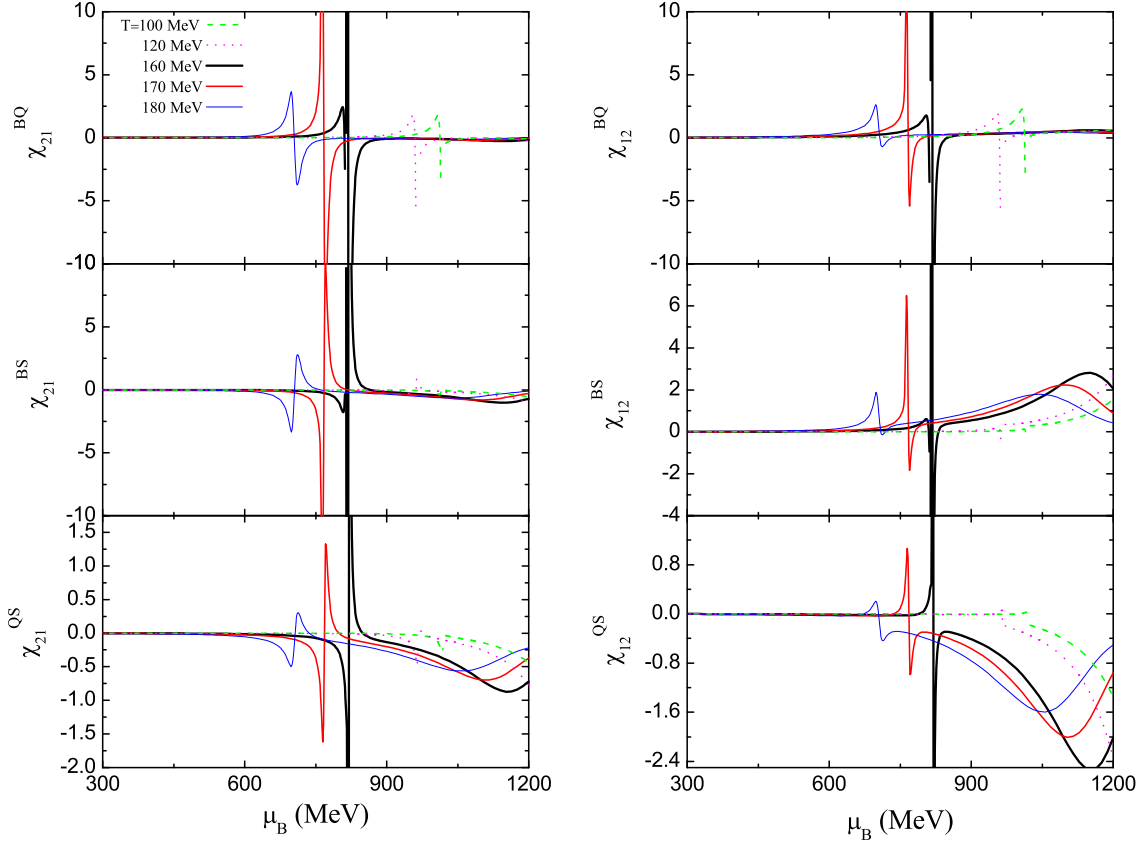}
\caption{(color online). Third-order correlations $\chi^{BQ}_{21}$
(top-left), $\chi^{BQ}_{12}$ (top-right), $\chi^{BS}_{21}$
(middle-left), $\chi^{BS}_{12}$ (middle-right), $\chi^{QS}_{21}$
(bottom-left), and $\chi^{QS}_{12}$ (bottom-right) as functions of
the baryon chemical potential $\mu_{B}$ ($\mu_{Q}=\mu_{S}=0$) with
several values of temperature in the PNJL model.}\label{f9}
\end{figure}

In Fig.~\ref{f9} we plot the third-order correlations
$\chi^{BQ}_{21}$, $\chi^{BQ}_{12}$, $\chi^{BS}_{21}$,
$\chi^{BS}_{12}$, $\chi^{QS}_{21}$, and $\chi^{QS}_{12}$ as
functions of the baryon chemical potential at several values of
temperature calculated in the PNJL model. It is seen that all these
third-order correlations change their signs at the chiral phase
transition, which are the same as the third-order fluctuations of
conserved charges. More concretely, $\chi^{BS}_{21}$ and
$\chi^{QS}_{21}$ change their signs from negative to positive with
the increase of the baryon chemical potential during the chiral
phase transition, while other correlations in Fig.~\ref{f9} changes
their signs in the opposite direction. Furthermore, one can observe
that the oscillating amplitudes of these third-order correlations
all increase rapidly when moving toward the QCD critical point and
diverge at the critical point.

\begin{figure}[!htb]
\includegraphics[scale=1.2]{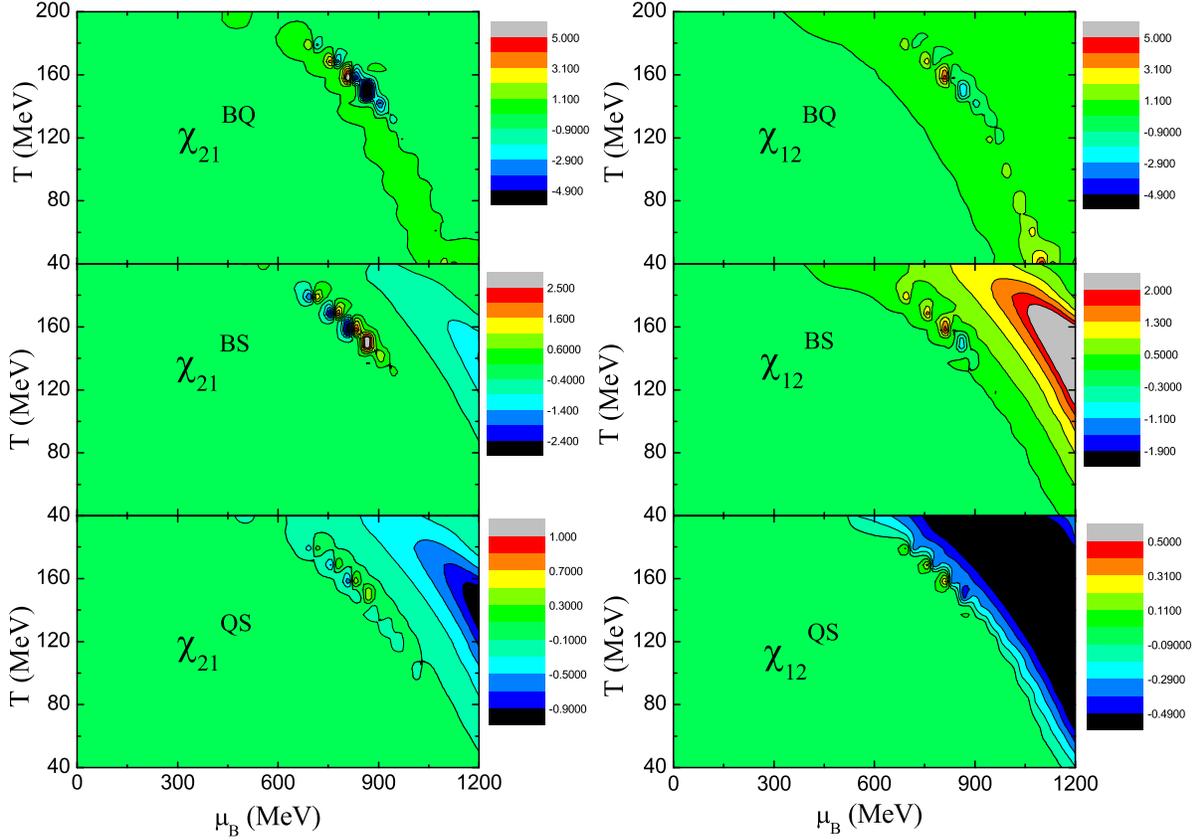}
\caption{(color online). Contour plots of the third-order
correlations $\chi^{BQ}_{21}$ (top-left), $\chi^{BQ}_{12}$
(top-right), $\chi^{BS}_{21}$ (middle-left), $\chi^{BS}_{12}$
(middle-right), $\chi^{QS}_{21}$ (bottom-left), and $\chi^{QS}_{12}$
(bottom-right) as functions of temperature $T$ and baryon chemical
potential $\mu_{B}$ ($\mu_{Q}=\mu_{S}=0$) in the PNJL
model.}\label{f10}
\end{figure}

Fig.~\ref{f10} shows the contour plots of the third-order
correlations $\chi^{BQ}_{21}$, $\chi^{BQ}_{12}$, $\chi^{BS}_{21}$,
$\chi^{BS}_{12}$, $\chi^{QS}_{21}$, and $\chi^{QS}_{12}$ as
functions of temperature and baryon chemical potential calculated in
the PNJL model. Comparing Fig.~\ref{f10} and Fig.~\ref{f8}, one can
easily notice that the QCD critical point in Fig.~\ref{f10} is much
more obvious than that in Fig.~\ref{f8}, which means that higher
order correlations are more sensitive to the critical behavior of
the QCD critical point than the quadratic correlations, and are
better to be used for exploring the critical behavior of the QCD
critical point in heavy ion collision experiments. As for the six
contour plots in Fig.~\ref{f10}, it is easily seen that the QCD
critical point in $\chi^{BQ}_{21}$, $\chi^{BS}_{21}$, and
$\chi^{QS}_{21}$ is more distinct than that in $\chi^{BQ}_{12}$,
$\chi^{BS}_{12}$, and $\chi^{QS}_{12}$.

\begin{figure}[!htb]
\includegraphics[scale=0.8]{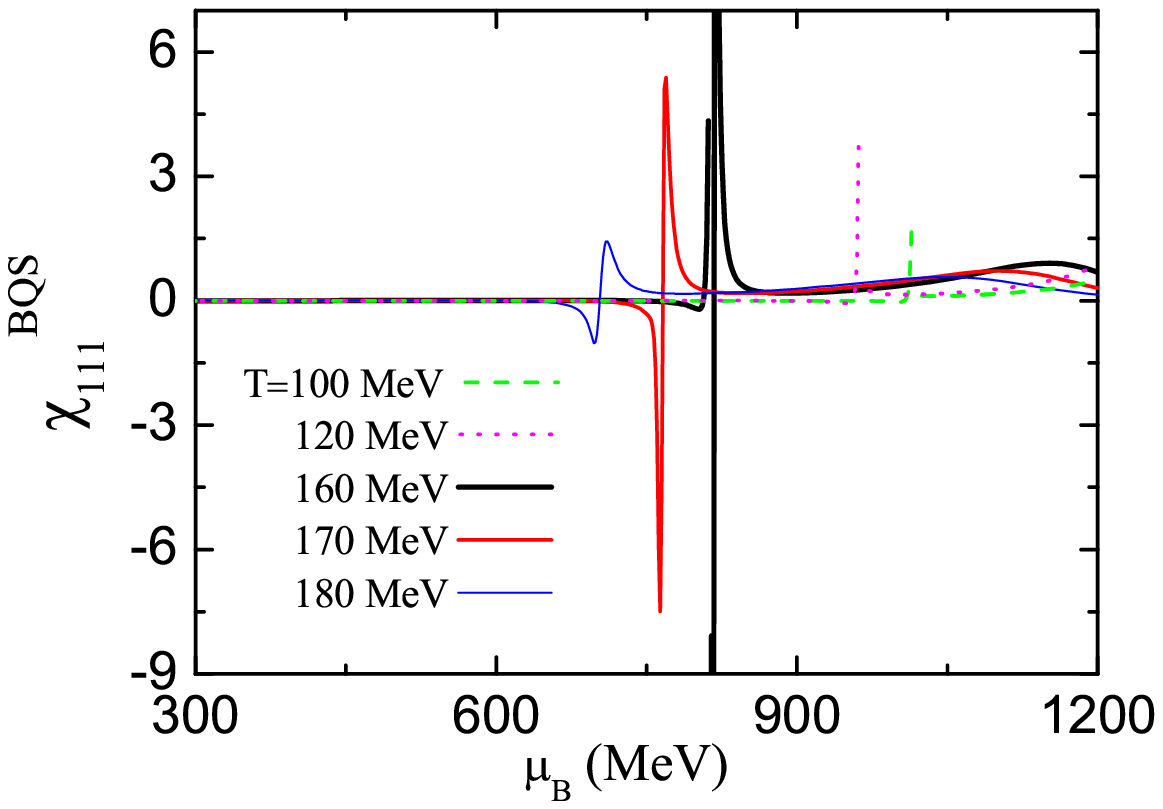}
\caption{(color online). Third-order correlation among baryon
number, electric charge, and strangeness, i.e., $\chi^{BQS}_{111}$
as a function of the baryon chemical potential $\mu_{B}$
($\mu_{Q}=\mu_{S}=0$) with several values of temperature in the PNJL
model.}\label{f11}
\end{figure}

\begin{figure}[!htb]
\includegraphics[scale=0.9]{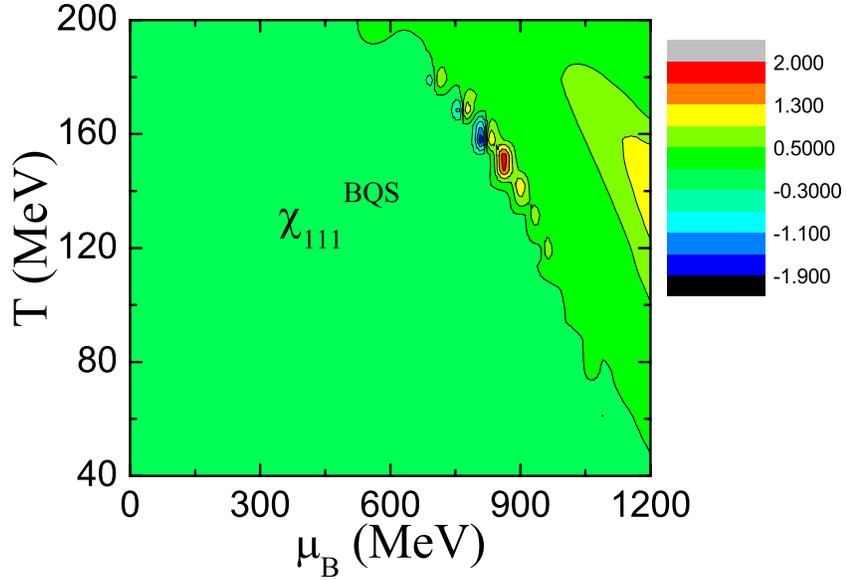}
\caption{(color online). Contour plot of the third-order correlation
$\chi^{BQS}_{111}$ as the function of temperature $T$ and baryon
chemical potential $\mu_{B}$ ($\mu_{Q}=\mu_{S}=0$) in the PNJL
model.}\label{f12}
\end{figure}

In Fig.~\ref{f11} we present the last third-order correlation
$\chi^{BQS}_{111}$, i.e., the correlation among the baryon number,
electric charge, and the strangeness, as a function of the baryon
chemical potential with several values of temperature calculated in
the PNJL model. We see that $\chi^{BQS}_{111}$ changes its sign from
negative to positive during the chiral phase transition and diverges
at the QCD critical point. Furthermore, it is noticed that only when
the thermodynamical system is near the chiral phase transition,
$\chi^{BQS}_{111}$ has nonvanishing value. We also show the
corresponding contour plot of the $\chi^{BQS}_{111}$ as the function
of $T$ and $\mu_{B}$ in Fig.~\ref{f12}, and it is seen that the QCD
critical point in the contour plot of $\chi^{BQS}_{111}$ is very
obvious. Therefore, $\chi^{BQS}_{111}$ is an ideal probe to search
for the QCD critical point in heavy ion collision experiments.

\section{Summary and Discussions}

We have studied the fluctuations and correlations of conserved
charges, i.e., the baryon number, the electric charge and the
strangeness, in the 2+1 flavor Polyakov--Nambu-Jona-Lasinio model at
finite temperature with nonzero baryon chemical potential. More
attentions have been paid on the studies of the non-monotonic
behavior of the fluctuations and correlations of conserved charges
near the QCD critical point. The fluctuations are calculated up to
the fourth-order and the correlations to the third-order.

It has been shown that the second-order fluctuations and
correlations have a peak or valley structure when the chiral phase
transition takes place with the increase of the baryon chemical
potential. As for the higher order fluctuations and correlations of
conserved charges, it has been seen that the third-order
fluctuations and correlations change their signs during the chiral
phase transition and the fourth-order fluctuations have two maximum
and one minimum. Furthermore, it has been noticed that the absolute
values of the extrema of the fluctuations and correlations at the
chiral phase transition increase rapidly when the thermodynamical
system moves toward the QCD critical point (in heavy ion collision
experiments, which means that the freeze-out point moves toward the
QCD critical point), and finally the fluctuations and correlations
of conserved charges diverge at the critical point.

We have also explicitly demonstrated the critical behavior by
depicting contour plots of the fluctuations and correlations of
conserved charges. In these contour plots, one can clearly figure
out the chiral phase transition line. Comparing with the
second-order fluctuations and correlations, we have found that
higher order cumulants, such as the third- and fourth-order
fluctuations and the third-order correlations discussed in this
paper, are more sensitive to the critical behavior of the QCD
critical point. Therefore, we arrive at the conclusion that the
higher order fluctuations and correlations of conserved charges are
superior to the second-order ones to be used to search for the
critical point in heavy ion collision experiments. Particularly, we
would like to address that among all the fluctuations and
correlations discussed in this paper, the numerical calculations
within the 2+1 flavor PNJL model indicate that $\chi^{BQ}_{21}$,
$\chi^{BS}_{21}$, $\chi^{QS}_{21}$, and $\chi^{BQS}_{111}$ are the
most valuable probes for exploring the QCD critical point.

\section*{Acknowledgements}

W. J. F. acknowledges financial support from China Postdoctoral
Science Foundation No. 20090460534. Y. L. W. is supported in part by
the National Science Foundation of China (NSFC) under the grant No.
10821504.

\end{document}